\documentclass[twocolumn,aps]{revtex4}
\usepackage{longtable,setspace}
\usepackage{graphicx,rotating,lscape}
\usepackage{color}
\usepackage{float,epsfig,amsmath,amssymb,euscript,color}
\usepackage{amsmath,amssymb,epsfig,capt-of,ifthen,calc}
\usepackage{float,euscript}
\usepackage{aecompl}
\usepackage{color}
\usepackage{soul}
\usepackage{multirow}
\usepackage{dcolumn}
\usepackage{longtable}
\usepackage{supertabular}
\usepackage{setspace}
\usepackage{ulem,enumitem}
\usepackage{bm}
\usepackage{hyperref}
\usepackage{natbib}
\usepackage{amsthm}

\definecolor{bleuclair}{rgb}{0.7, 0.7, 1.0}
\definecolor{rosepale}{rgb}{1.0, 0.7, 1.0}

\begin{document}

\title{Searching for crystal-ice domains in amorphous ices}
\author{Fausto Martelli$^{1,2,*}$, Nicolas Giovambattista$^{3,4}$, Salvatore Torquato$^{2}$, Roberto Car$^{2}$}
\affiliation{%
$^{1}$IBM Research, Hartree Centre, Daresbury, WA4 4AD, United Kingdom \\
$^{2}$Department of Chemistry, Princeton University, Princeton, New Jersey 08544, USA \\
$^{3}$Department of Physics, Brooklyn College of the City University of New York, Brooklyn, New York 11210, USA \\
$^{4}$The Graduate Center of the City University of New York, New York, New York 10016, USA \\
$^{*}$Corresponding author: fausto.martelli@ibm.com \\
}

\begin{abstract}
We employ classical molecular dynamics simulations to investigate the molecular-level structure of water during the isothermal 
compression of hexagonal ice (I$h$) and low-density amorphous (LDA) ice at low temperatures. In both cases, the system transforms 
to high-density amorphous ice (HDA) via a first-order-like phase transition. 
We employ a sensitive local order metric (LOM) [Martelli \textit{et. al.}, \textit{Phys. Rev. B}, \textbf{97}, 064105 (2018)],
that can discriminate among different crystalline and non crystalline ice structures and is based on 
the positions of the oxygen atoms in the first and/or second hydration shell. 
Our results confirm that LDA and HDA are indeed amorphous, i.e., they lack of polydispersed ice 
domains. Interestingly, HDA contains a small number of domains
that are reminiscent of the unit cell of ice IV, although the hydrogen-bond network (HBN) of these domains differ 
from the HBN of ice IV. The presence of ice IV-like domains provides some support to the hypothesis that HDA 
could be the result of a detour on the HBN rearrangement along the I$h$-to-ice IV pressure 
induced transformation.
Both nonequilibrium LDA-to-HDA and I$h$-to-HDA transformations are two-steps processes where a small distortion of 
the HBN first occurs at low pressures and then, a sudden, extensive re-arrangement of hydrogen
bonds at the corresponding transformation pressure follows. 
Interestingly, the I$h$-to-HDA and LDA-to-HDA transformations occur when LDA and I$h$ have
similar local order, as quantified by the site-averaged LOMs. Since I$h$ has a perfect tetrahedral HBN, while LDA does not, 
it follows that higher pressures are needed to transform I$h$ into HDA than that for the conversion of LDA to HDA.
In correspondence with both first-order-like phase transitions, the samples are composed of a large 
HDA cluster that percolates within the I$h$/LDA samples. \\
Our results shed light on the debated structural properties of amorphous ices and indicate that the kinetics 
of the I$h$-to-HDA and LDA-to-HDA transformations require an in depth inspection of the underlying HBN. Such investigation is 
currently ongoing.
\end{abstract}

\maketitle

\section{Introduction}
At deeply supercooled conditions, water exhibits polyamorphism, i.e., it exists in more
than one amorphous solid state. The most common forms
of glassy water are the low-density amorphous (LDA) and the high-density amorphous (HDA) ice~\cite{debenedetti_2003,angell_2004,giovanbattista_2006,mishima_1998,loerting_the_2015,aman_colloquium_2016}. 
LDA is likely the most abundant form of ice in the universe and can be produced, for example, by 
rapid quenching of liquid water at atmospheric pressure~\cite{meyer_new_1985}.
The LDA that is thought to exist in space forms by condensation of water from the gas phase onto cold 
surfaces~\cite{burton_nature_1935,jenniskens_science_1935}.
HDA can be produced, for example, by pressure induced amorphization (PIA) of hexagonal ice (I$h$) or by isothermal compression of 
LDA~\cite{mishima_reversible_1993,mishima_melting_1984,andersson_glass_2011,loerting_amorphous_2006}.
Remarkably, LDA and HDA can be inter-converted by isothermal compression/decompression at 
T$=130-140$ K~\cite{mishima_reversible_1993,winkel_structural,winkel_water} and
by isobaric heating at different pressures~\cite{winkel_structural,winkel_equilibrated_2011,lyapin_2011}. A third form of 
glassy water, a very high-density amorphous ice (VHDA), has been identified at very high pressures~\cite{loerting_second_2001}.
We note that experiments show that LDA and HDA can be separated into subfamilies of amorphous structures, e.g., LDA$_I$ and
LDA$_{II}$ for LDA~\cite{winkel_2009}, and unannealed HDA (uHDA) and expanded HDA (eHDA) for the case of 
HDA~\cite{nelmes_2006,loerting_how,loerting_the_2015,aman_colloquium_2016}. However, this distinction is less clear in computer simulations 
and, therefore, we refer to glassy water as either LDA or HDA. \par
At small length scales, LDA and HDA are structurally very different. LDA has well-separated first and second hydration shells, with nearest-neighbors arranged in a tetrahedral-like 
local structure. In this regard, the structure of LDA is reminiscent of the local structure of I$h$.
By contrast, HDA has interstitial molecules
populating the space between the first and the second hydration shells, thus acquiring distorted
local configurations similar to liquid water configurations at ambient 
conditions~\cite{santra_local,tulk_structural_2002,finney_structures_2002,winkel_equilibrated_2011,soper_structures_2000,nicolas_pressure_2015}.
LDA and HDA also differ in terms of the hydrogen-bond network (HBN). The HBN of LDA is dominated by $5$-, $6$-, 
and $7$-fold rings, in contrast to HDA, whose HBN includes a significant fraction of longer member rings to 
accommodate the larger density of the system ($\sim20-25\%$ larger than LDA)~\cite{martonak}.
Despite these local structural differences, at large length scales the two glass forms are nearly hyperuniform, i.e., they possess 
similar degree of suppression of large-scale density fluctuations, indicating that they should possess similar 
large-scale structures and large-scale translational order~\cite{martelli_hyperuniformity}. \par
LDA and HDA were discovered more that 30 years ago~\cite{debenedetti_2003}. Yet, the nature of LDA and HDA, and the associated 
LDA-to-HDA first-order-like phase transition, remain highly debated.  
It has been suggested that LDA and HDA are thermodynamically connected with the liquid, e.g., by isobaric 
cooling/heating~\cite{mishima_reversible_1993}. Elsewhere, 
HDA was interpreted to be a collapsed HBN of water molecules, unrelated to the liquid 
state~\cite{tse_mechanism,johari_2000,johari_2004}. It has also been suggested that HDA may contain nanometer-scale ice domains.  
After all, HDA transforms to ice IV at very high pressure~\cite{salzmann_2002,salzmann_2004} and it can be 
formed by compression of I$h$. Unfortunately,   
experimental studies that focus on the structure of LDA and HDA at intermediate lengths scales are 
challenging~\cite{aman_colloquium_2016}. Computer simulation studies that describe the structure of LDA and HDA, 
especially the search for the presence of ice-like domains, are rare. \par
In this article, we perform out-of-equilibrium classical MD simulations to study the structural order 
during the I$h$-to-HDA and LDA-to-HDA 
transformations; in particular, we look for traces of crystalline domains during these process.
We will refer to the HDA produced upon compression of I$h$ as HDA$_{Ih}$, and to the HDA 
obtained upon compression of LDA as HDA$_{LDA}$. 
We probe the short- and intermediate-range order during these transformations using a recently 
developed local order metric (LOM). The LOM measures the degree of order present in the neighborhood of an atomic or molecular site
in a condensed medium~\cite{martelli_LOM}. The LOM is endowed with a high-resolving 
power~\cite{martelli_LOM,martelli_confined,santra_BNT} 
and allows one to look for specific ordered domains defined by the location of selected atoms (e.g., water oxygens) 
in a given reference structure. Typically, the reference structure is taken to be the local structure of a perfect crystalline phase.\par
We have looked for signatures of ices I$h$, cubic (I$c$), II, III, IV, V, VI, VII, and VIII in LDA and HDA.
According to our analysis, both amorphous ices lack polydispersed ordered crystalline domains. However,
we find that the oxygens of few water molecules in HDA are arranged as in the unit cell of ice IV.
This observation provides support to the picture proposed in Ref.~\cite{shephard_HDA}, 
where the collapse of the HBN of I$h$ occurring upon isothermal compression does not lead 
to ice IV, but to HDA, hence considered a 'derailed' state along the I$h$-to-ice IV pathway.
Our observation is also consistent with the transformation of
HDA to ice IV reported in experiments at very high pressures~\cite{salzmann_2002,salzmann_2004}.
On the other hand, we also find that the HBN connecting the water molecules in
these ice IV-like domains in HDA {\it differs} from the HBN in the unit cell of ice IV. \par
Our analysis indicateis that the I$h$-to-HDA transformation is a two-step
process in which, first, compression causes a continuous
distortion of the ordered HBN of I$h$. This continuous distortion of the HBN is then followed by a sudden 
extensive rearrangement of the HBN that occurs in correspondence with the I$h$-to-HDA
first-order-like phase transition.
A similar two-step process occurs during the LDA-to-HDA transformation. In this case, however, although LDA and I$h$ acquire similar
tetrahedral configurations, the LDA-to-HDA transformation is milder than the I$h$-to-HDA transformation, and it occurs at lower
pressures. We notice that the second hydration shell in I$h$ is well ordered.
Conversely, the second hydration shell in LDA describes an 'open cage' with a disordered HBN and, therefore, is less rigid. Hence, 
we relate the lower transformation pressure in LDA, relative to I$h$, to the lower rigidity of the second hydration shell
in LDA compared to I$h$. We complement our analysis by performing a clustering analysis of local environments at different 
pressures and show that both LDA-to-HDA and I$h$-to-HDA transformations are reminiscent to spinodal decompositions, without 
nucleation and growth of HDA within LDA/I$h$. \par
The article is organized as follows. In Section~\ref{LOM}, we provide a brief definition of the LOM employed in this work.  
In Secs.~\ref{short} and ~\ref{inter}, we discuss the short- and the intermediate-range order, respectively,
in both LDA and HDA. In Section~\ref{cluster}, we discuss
the spatial aggregation (clusters) of LDA-, HDA-, and I$h$-like molecules during the LDA-to-HDA and I$h$-to-HDA transformations.
Conclusions and final remarks are presented in Sec.~\ref{Conclusions}.
\section{The local order metric}\label{LOM}
The local environment of an atomic site $j$ in a snapshot of a molecular dynamics or
Monte Carlo simulation defines a local pattern formed by $M$ neighboring sites. Typically
these include the first and/or the second neighbors of the site $j$. There are $N$ local patterns,
one for each atomic site $j$ in the system. 
The local reference structure is the set of the same $M$ neighboring sites
in an ideal lattice of choice, the spatial scale of which is fixed by setting its nearest
neighbor distance equal to $d$, the average equilibrium value in the system of interest. 
For a given orientation of the reference structure and a given permutation $\mathcal{P}$ of the pattern indices,
we define the LOM $S(j)$ as the maximum overlap between pattern and reference structure in the $j$
neighborhood by:
\begin{equation}
  S(j)=\max_{\theta,\phi,\psi;\mathcal{P}}\left\{\prod_{i=1}^{M}\exp\left(-\frac{\left| \mathbf{P}^{j}_{i\mathcal{P}}-\mathbf{A}^{j}\mathbf{R}_{i}^{j}\right|^2}{2\sigma^{2}M}\right)\right\}
  \label{eq:Eq1}
\end{equation}
Where $\theta, \phi$ and $\psi$ are Euler angles, $\mathbf{P}^{j}_{i\mathcal{P}}$ and $\mathbf{R}_{i}^{j}$ are the pattern and the reference
position vectors in the laboratory frame of the $M$ neighbors of site $j$, respectively, and $\mathbf{A}^{j}$ is an arbitrary 
rotation matrix about the pattern centroid.
The parameter $\sigma$ controls the spread of the Gaussian functions. 
The LOM satisfies the inequalities $0 \lesssim S(j) \leq 1$. The two limits
correspond, respectively, to a completely disordered local pattern ($S(j)\rightarrow 0$) and to an
ordered local pattern matching perfectly the reference ($S(j)\rightarrow 1$). 
We also define a global order parameters based on $S(j)$, as the average score function $S$: 
\begin{equation}
  S=\frac{1}{N}\sum_{j=1}^{N}S(j)
  \label{eq:Eq3}
\end{equation}
To improve statistics, the score functions are also averaged over $10$ independent (LDA-to-HDA and I$h$-to-HDA) trajectories.
\section{Results}\label{Results}
Our study is based on classical molecular dynamics simulations 
of a system of $N=8192$ water molecules described by the classical TIP4P/2005 interaction potential~\cite{abascal_tip4p2005}.
This water model is able to reproduce relatively well the structures of LDA and HDA at low 
temperatures~\cite{nicolas_pressure_2015}. Computer simulations details and a description of the protocols 
employed in this 
work are provided in Ref.~\cite{engstler_2017}. Briefly, we prepare
LDA by cooling an equilibrium liquid from $240$ K to $80$ K, with a cooling rate of $1$ K/ns. HDA is obtained by isothermal 
compression of I$h$ and LDA at $T=80$ K. 
During the compression of I$h$ and LDA, the pressure is increased from ambient to $3.0$ GPa at a pace 
of $0.01$ GPa/ns. For the water model considered, at the present cooling and compression rates,
the I$h$-to-HDA transformation occurs at $p=1.35$ GPa, while the LDA-to-HDA transformation occurs at $0.83\lesssim p\lesssim0.93$ 
GPa~\cite{engstler_2017}. All results reported in this work are averaged over $10$ independent trajectories. \par
\subsection{Short-range order: local tetrahedrality}\label{short}
\begin{figure}
 \centering
    \includegraphics[scale=.30]{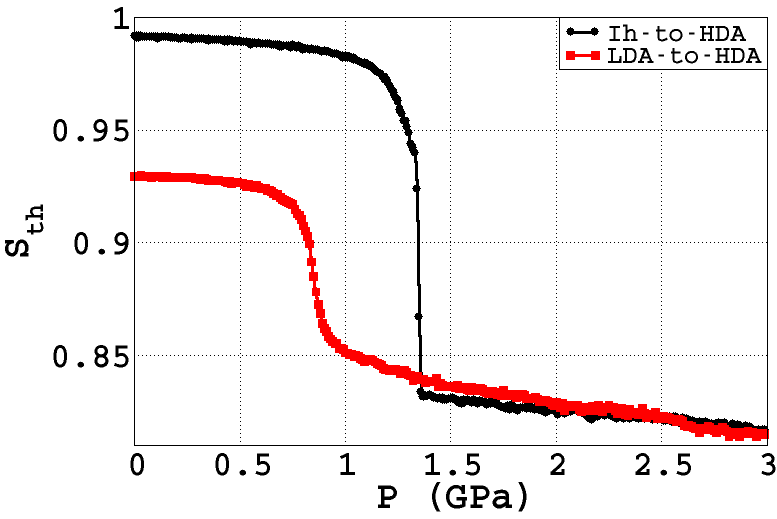}
    \caption{Local tetrahedrality during the I$h$-to-HDA (black circles) and LDA-to-HDA (red squares) transformations. 
            $S_{th}$ is the score function defined using a regular tetrahedron as a reference structure. In both transformations, 
            compression at low pressures leads to a continuous decrease in local tetrahedrality. The sudden jump in $S_{th}$ at 
            $p\sim0.83$ GPa (LDA) and $p\sim1.36$ GPa (I$h$) coincides with the sharp transformation of the system to 
            HDA~\cite{engstler_2017}.}
 \label{fig:FigSth}
\end{figure}

In this section, we discuss the local structure of water during the LDA-to-HDA and I$h$-to-HDA transformation at 
the level of the first hydration shell. Experiments and computer simulations indicate that the local structure of I$h$ and LDA 
is tetrahedral, i.e., a given molecule is located at the center of a tetrahedron and its four nearest-neighbors are roughly 
located at the corner of such a tetrahedron. Therefore, in order to probe the short-range order of water during the LDA-to-HDA and 
I$h$-to-HDA transformations, we consider a LOM defined using, as a reference structure, a regular tetrahedron. The resulting 
score function, $S_{th}$, is shown in Fig.~\ref{fig:FigSth} for the compression of I$h$ (black circles) and LDA (red squares).
At low applied pressures, both I$h$ and LDA acquire high values of $S_{th}$, reflecting the nearly perfect regular tetrahedrality of I$h$ and LDA.
Upon compressing the samples, $S_{th}$ slightly decreases as the local tetrahedral structures get more and more distorted.
The transformation of LDA and I$h$ to HDA occur, respectively, at $0.83\lesssim p\lesssim 0.93$ GPa and 
$1.35\lesssim p \lesssim 1.36$ GPa. Accordingly, Fig.~\ref{fig:FigSth} shows that, at these pressures, $S_{th}$ 
decays sharply; the change in $S_{th}$ being more abrupt for the case of I$h$ than for LDA.\\
The behavior of $S_{th}$ in Fig.~\ref{fig:FigSth} is consistent with the evolution of the tetrahedral order 
parameter $q$~\cite{chau_a_new_1998,errington_relationship_2001} during the I$h$-to-HDA and LDA-to-HDA transformations reported
in Ref.~\cite{engstler_2017}. In particular, 
it was found that both I$h$-to-HDA and LDA-to-HDA transformations occur when 
the tetrahedral order parameter reaches a specific critical value, $q\sim0.32$. Similarly, Fig.~\ref{fig:FigSth} indicates 
that both transformations occur when $S_{th}$ reaches the critical value
$S_{th}\sim0.91$ and $\sim0.94$ in LDA and I$h$, respectively.
This strongly suggests that the transformation of LDA and I$h$ to HDA have a common origin, the rearrangement of the 
HBN~\cite{martelli_prep}that occurs at $S_{th}\sim0.91-0.94$. This also explains why the transformation pressure is higher for I$h$ 
than for LDA. Specifically, 
the HBs in I$h$ are more linear (and thus, stronger) than in LDA and, hence, in order to distort the HBN to the point that 
$S_{th}\sim0.91-0.94$, one needs to apply larger pressures to I$h$ than to LDA. \par

\begin{figure}
 \centering
    \includegraphics[scale=.40]{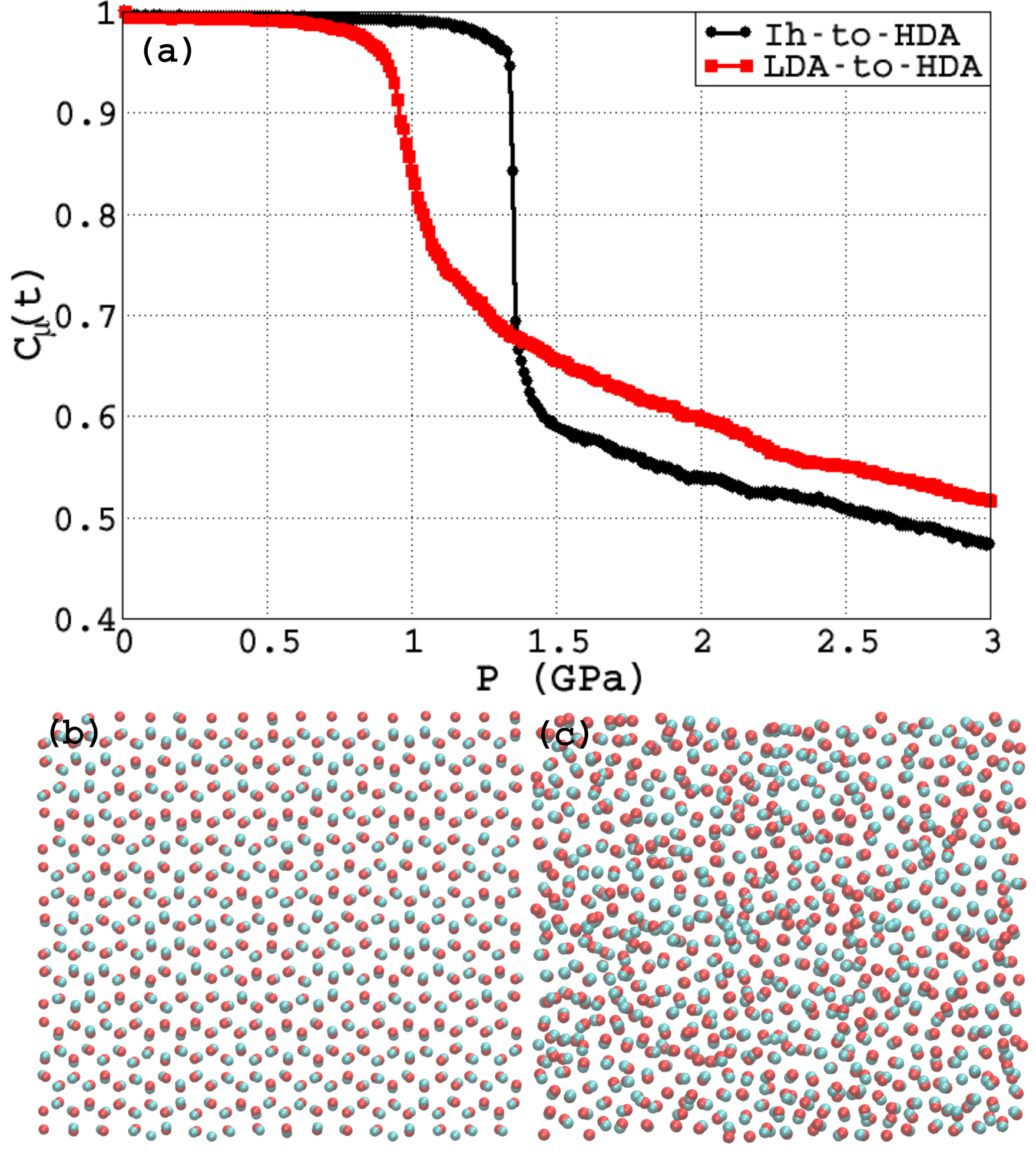}
    \caption{(a) Water dipole moment correlation function $C_{\mu}(p)$ as a function of pressure during the
             I$h$-to-HDA (black circles) and LDA-to-HDA transformations (red squares). (b) Snapshot of
             a $4$ \AA-thick slice of I$h$ at $p = 0$ GPa. (c) Snapshot of a $4$ \AA-thick slice
             of HDA at $p=3.0$ GPa. Red spheres represent the oxygen 
             atoms, and the cyan spheres indicate the position of the corresponding dipole end-point.}
 \label{fig:dipole}
\end{figure}

To shed light into the microscopic origin of the LDA-to-HDA and I$h$-to-HDA transformations, we also study the
normalized molecular dipole correlation function, 
$C_{\mu}(p)=\left<\mu(p)\cdot\mu(0)\right>$, 
where $\mu(0)$ is the molecular dipole 
vector at pressure $p=0$ GPa, $\mu(p)$ is the molecular dipole vector at pressure $p$ and $<\cdot>$ indicates  
a average over all molecules in the system.
Figure~\ref{fig:dipole} (a) shows the profile of $C_{\mu}(p)$ for the I$h$-to-HDA transformation (black circles)
and for the LDA-to-HDA transformation (red squares). At low pressures, shows only a mild decay upon compression. 
In correspondence with both transition pressures, $C_{\mu}(p)$ shows a marked 
drop, indicating that water molecules rotate causing a rearrangement in the HBN. In particular, the drop in $C_{\mu}(p)$ for the 
transformation of I$h$ is larger than the drop in $C_{\mu}(p)$ for the transformation of LDA.
As an example, we include in Fig.~\ref{fig:dipole} (b) and (c) two snapshots
of a $4$\AA-thick slice of I$h$ at $p=0$ GPa and of HDA at $p=3$ GPa. The red spheres correspond to oxygen atoms, while 
the cyan spheres correspond to the positions of the end-point of the molecular dipoles. One can appreciate how the ordered
pattern in the dipoles distribution on I$h$ is broken in HDA.\\
\subsection{Intermediate-range order: searching for ice I$h$, I$c$ and II-VIII}\label{inter}
\begin{figure}
 \centering
    \includegraphics[scale=.33]{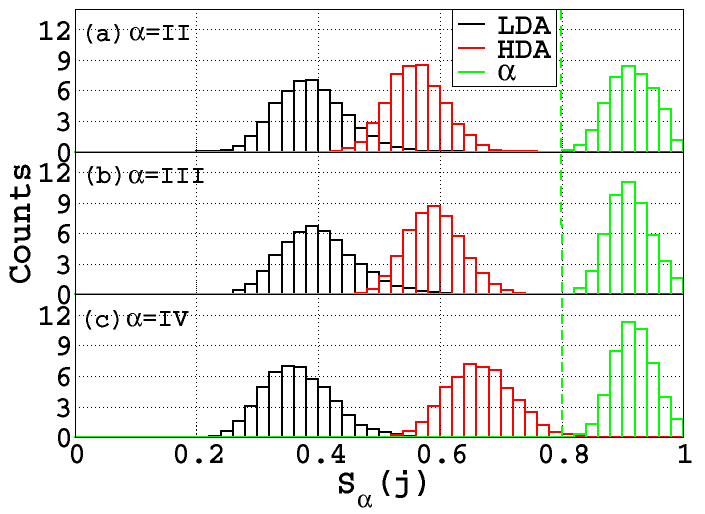}
    \caption{Histogram of the LOM $S_{\alpha}(j)$, eq.~\ref{eq:Eq1} calculated using, as a reference, the unit 
            cells of ice (a) $\alpha=$II, (b) $\alpha=$III and (c) $\alpha=$IV for LDA (black)
            and for HDA$_{Ih}$ (red) at p$=0.01$ GPa and at p$=3.0$ GPa, respectively. The green distrubutions represent the histograms
            for the corresponding bulk ices computed at the following thermodynamic conditions~\cite{conde_2013}: $T=210$ K and 
            $p=4$ GPa for ice II, and $T=250$ K and $p=3$ GPa for ice III and IV. The green line emphasizes the 
            value $S=0.8$ used as a cutoff to identify ice-like environments.}
 \label{fig:hist}
\end{figure}

In this section, we focus on the local structure of water at intermediate length scales. Specifically, we study the structure 
of water during the LDA-to-HDA and I$h$-to-HDA transformations at the level of the second hydration shell. This allows us to 
compare the structure of LDA and HDA relative to the structure of I$h$ as well as intermediate and high-pressure ices.
Water can acquire more than 18 different crystalline forms~\cite{bartles_2012}. However, at the temperature ($T=80$~K) and pressures 
($0\leq p \leq 3$~GPa) studied here, the system can only visit the stability regions of the phase diagram of ice corresponding to I$h$ 
and I$c$, as well as ices II, III, IV, V, VI, VII and VIII. Accordingly, we focus on the score function $S_{\alpha}$ associated to ice
$\alpha=$ I$h$, I$c$, II, III, IV, V, VI, VII and VIII, and defined using, as a reference structure, the first and/or
second hydration shell of the corresponding ice.
\begin{center}
 \begin{table*}[th]
   \begin{tabular}{ |c|c|c|c|c|c|c| }
     \hline
      Ice $\alpha$  & M  & Shells      & $N_{cryst}$ (LDA) & $S_{\alpha}$ (LDA) & $N_{cryst}$ (HDA) & $S_{\alpha}$ (HDA) \\ \hline
      I$h$ & 12 & 2nd         & 0 & 0.416$\pm$0.085 & 0 & 0.581$\pm$0.039  \\
      I$c$ & 12 & 2nd         & 0 & 0.331$\pm$0.081 & 0 & 0.557$\pm$0.039  \\
      II   & 11 & 1st and 2nd & 0 & 0.389$\pm$0.054 & 0 & 0.562$\pm$0.049  \\
      III  & 11 & 1st and 2nd & 0 & 0.403$\pm$0.015 & 0 & 0.592$\pm$0.054  \\ 
      IV   & 13 & 1st and 2nd & 0 & 0.368$\pm$0.069 & 114 & 0.665$\pm$0.054 \\ 
      V    & 14 & 1st and 2nd & 0 & 0.327$\pm$0.021 & 0 & 0.46$1\pm$0.054 \\ 
      VI   & 11 & 1st and 2nd & 0 & 0.204$\pm$0.033 & 0 & 0.404$\pm$0.054 \\ 
      VII  & 19 & 1st and 2nd & 0 & 0.312$\pm$0.028 & 0 & 0.469$\pm$0.054  \\ 
      VIII & 18 & 1st and 2nd & 0 & 0.296$\pm$0.020 & 0 & 0.438$\pm$0.054  \\ 
     \hline
   \end{tabular}
 \caption{Score function $S_{\alpha}$ (averaged over $10$ independent trajectories) for LDA at $p=0.1$ MPa and HDA at $p=3.0$ GPa.  
          $S_\alpha$ is the score function defined using, as a reference structure, the first and/or second shell of 
          ice $\alpha=$ I$h$, I$c$, II-VIII. There are $M$ water oxygens in these reference structures.
          Included are the number of crystallites $N_{cryst}$ found in LDA and HDA for the ices studied (see text).}
 \label{tb:Table1}
 \end{table*}
\end{center}
Table~\ref{tb:Table1} shows the value of the score functions $S_{\alpha}$ for
the cases of LDA at $p=0.1$~MPa, and HDA at $p=3.0$~GPa. In the case of HDA, 
we obtain similar values of $S_\alpha$ for HDA${_{LDA}}$ and HDA${_{Ih}}$. 
Also included in Table~\ref{tb:Table1} are the number of crystallites found in LDA and HDA. For a given ice $\alpha$, 
we define a crystallite as a water oxygen $j$ for which the LOM $S_\alpha(j) > S_0$, plus its M neighbors used
in the reference structure of the corresponding LOM. $S_0$ is a cutoff 
reference value. In this work, we chose $S_0=0.8$ because we find that, for all ices studied, $S_\alpha(j)>0.8$.
As an example, we include in Fig.~\ref{fig:hist}(a)-(c) the probability distribution of $S_\alpha(j)$ in LDA (black) and HDA
(red), for the cases $\alpha=$ II, III, IV. In Fig.~\ref{fig:hist}(a)-(c) we also report the distributions of $S_{\alpha}$
obtained in samples of ice II, III and IV computed in their thermodynamic stability regions (green)~\cite{conde_2013}.
As shown in Table~\ref{tb:Table1}, these are among the ices with larger values of the score function 
$S_\alpha$ in HDA. \\
There are two main points that follow from our MD simulations. On one hand, (i) LDA and HDA
do not contain any crystalline domain, i.e., $S_\alpha(j)<0.8$ for all atoms in the systems (in the case of HDA, 
this holds for HDA prepared by compression of ice I$h$ and LDA). 
On the other hand, (ii) there are a few spread traces of ice IV in HDA. 
Specifically, the tail of the distribution of the LOM $S_{IV}(j)$ for HDA clearly extends to values larger than $0.8$ (Fig.~\ref{fig:hist} (c)), which is 
not the case of other ices (see, e.g, the cases of ice II and
III in Figs.\ref{fig:hist} (a) and \ref{fig:hist} (b)). In particular, the distribution 
of $S_{IV}(j)$ in HDA partially overlaps with the corresponding distribution in a pure sample of ice IV.
We stress that the number of ice IV-like domains is rather small; there are only 114 ice IV-like 
centers in HDA at $p=3.0$ GPa for the case $S_0=0.8$. Such molecules have a LOM 
$S_{IV}(j)\sim S_0=0.8$, indicating highly distorted environments. 
The absence of crystallites of ice I$h$, I$c$ and II-VIII and the very small amount of ice IV-like crystallites 
enable us to conclude that HDA 
and LDA are indeed true amorphous structures. Our conclusion is strengthened further by inspecting the HBN of the ice IV-like
crystallites, as discussed in the next Section.
\subsubsection{Score function based on ice IV}\label{sbased}
\begin{figure}
 \centering
    \includegraphics[scale=.30]{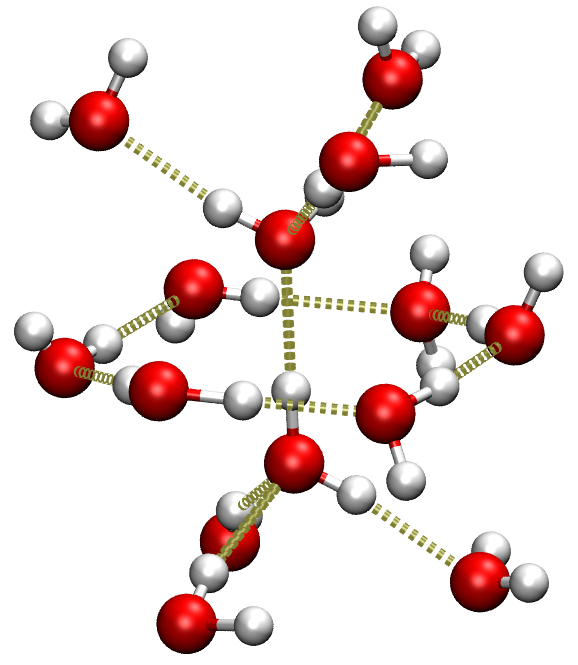}
    \caption{Unit cell of ice IV. The dashed lines represent the hydrogen bonds.}
    \label{fig:ice4}
\end{figure}

The presence of ice IV-like molecules in HDA is consistent with the crystallization of HDA into ice IV reported in 
high-pressure experiments~\cite{salzmann_2002,salzmann_2004}. 
Our results also provide support to the hypothesis that HDA may be an intermediate, 'derailed' glassy state formed during the 
I$h$-to-ice IV transformation~\cite{shephard_HDA}. However, we note that the HBN of the ice IV domains found in HDA 
differ from the HBN in ice IV. Specifically, the basic structure of ice IV 
is shown in Fig.~\ref{fig:ice4} and it is characterized by a water hexagon pierced by a donor-acceptor hydrogen bond.
This donor-acceptor HB is the origin of the interpenetrating HBN that characterizes ice IV~\cite{engelhardt_1981}.
We have inspected the topology of the HBN of all the $114$ ice IV-like domains in HDA at $p=3$ GPa
and found no indication of the donor-acceptor HB shown in Fig.~\ref{fig:ice4}. Hence, the ice IV-like domains 
should not be interpreted as nanoscopic crystals, 
but as disordered local domains in which the second shell of neighbors of the central molecule resembles 
a distorted second hydration shell of ice IV.
Indeed, network interpenetration constrains severely the configurational 
disorder of water's HBN and is discoraged by entropy. Thus, network interpenetration does not usually occur in disordered 
structures, such as the amorphous ices. Interestingly, when network interpenetration occurs, HDA does transform into ice IV,
as experimentally observed~\cite{salzmann_2002,salzmann_2003}.
It follows from our results that, although HDA may be considered as a 'derailed' state along the Ih-to-IV transformation, 
the microscopic mechanisms involved in this transformation requires further investigations~\cite{shephard_HDA}.\par
In order to delve deeper into the details of the ice-IV character of HDA, we show in Fig.~\ref{fig:scores} (a) 
the pressure-dependence of 
$S_{IV}$ during the I$h$-to-HDA transformation. The $S_{IV}$ is defined using, as a reference,
the unit cell of ice IV shown in Fig.~\ref{fig:scores} (a). For comparison, we also include 
$S_{IV}$ during the LDA-to-HDA transformation. Along both transformations, the ice IV-character of
both I$h$ and LDA increases smoothly with pressure and, in correspondence with both non-equilibrium transformations, 
$S_{IV}$ shows a sudden increase, from $S_{IV}=0.46-0.47$ to $S_{IV}= 0.64-0.66$. 
Interestingly, as for the sort-range $S_{th}$, both I$h$ and LDA need to acquire a similar value of $S_{IV}$ before being 
converted to HDA, i.e., $S_{IV}~\sim0.52$ at the corresponding transition pressures. This observation further explains why the 
transformation pressure is higher for I$h$ than for LDA, and indicates that such difference is an effect that extends also beyond
the first hydration shell. We also note that the continuous increment of the 
ice IV-character with the increasing pressure in HDA is a further indication of the HDA-to-ice IV transformation that may occur at 
higher pressures~\cite{salzmann_2002,salzmann_2004}.
\subsubsection{Score function based on I$h$}
\begin{figure}
 \centering
    \includegraphics[scale=.42]{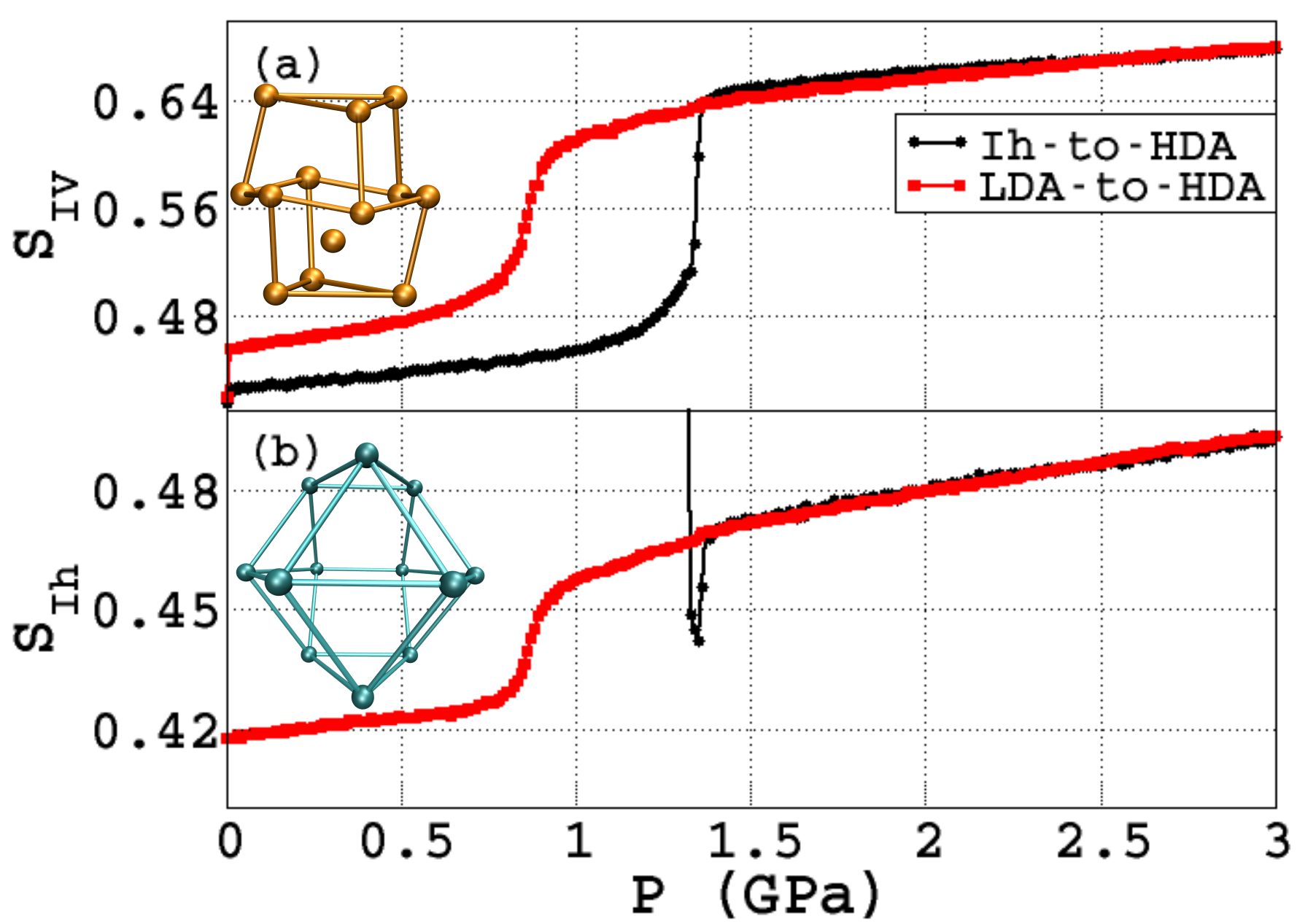}
    \caption{(a): Score function $S_{IV}$ during the I$h$-to-HDA (black) and the LDA-to-HDA (red) transformation.
             The snapshot represents the reference structure used to compute $S_{IV}$. (b) Score function 
             $S_{Ih}$ during the I$h$-to-HDA (black) and the LDA-to-HDA (red) transformation. The snapshot depicts the reference
             structure used in the definition of $S_{Ih}$.}
 \label{fig:scores}
\end{figure}

Considering that LDA and I$h$ have a similar degree of tetrahedrality (see Fig.~\ref{fig:FigSth}), one may wonder how 
similar LDA and I$h$ are at the level of the second hydration shell. To address this issue, we study S$_{Ih}$, 
i.e. the score function defined using, as a reference structure, the second hydration shell of ice I$h$ (see snapshot of 
Fig.~\ref{fig:scores} (b)). Figure~\ref{fig:scores} (b) shows $S_{Ih}$ during the I$h$-to-HDA (black line) 
and LDA-to-HDA (red line) 
transformations. The values of $S_{Ih}$ for I$h$ at low pressures are not shown because they are, as one would expect, 
close to $1$. Remarkably, in correspondence with 
the I$h$-to-HDA first-order-like phase transition, $S_{Ih}$ acquires a minimum that could be interpreted as the limit of 
mechanical stability for I$h$.
Further compression results in a continuous but mild increase in $S_{Ih}$ within the HDA$_{Ih}$ state. 
We note that $S_{Ih}$ is practically the same for both HDA$_{LDA}$ and HDA$_{Ih}$,
indicating that the average structure of HDA is independent of the recipe followed to prepare HDA. \par
\begin{figure}
 \centering
    \includegraphics[scale=.29]{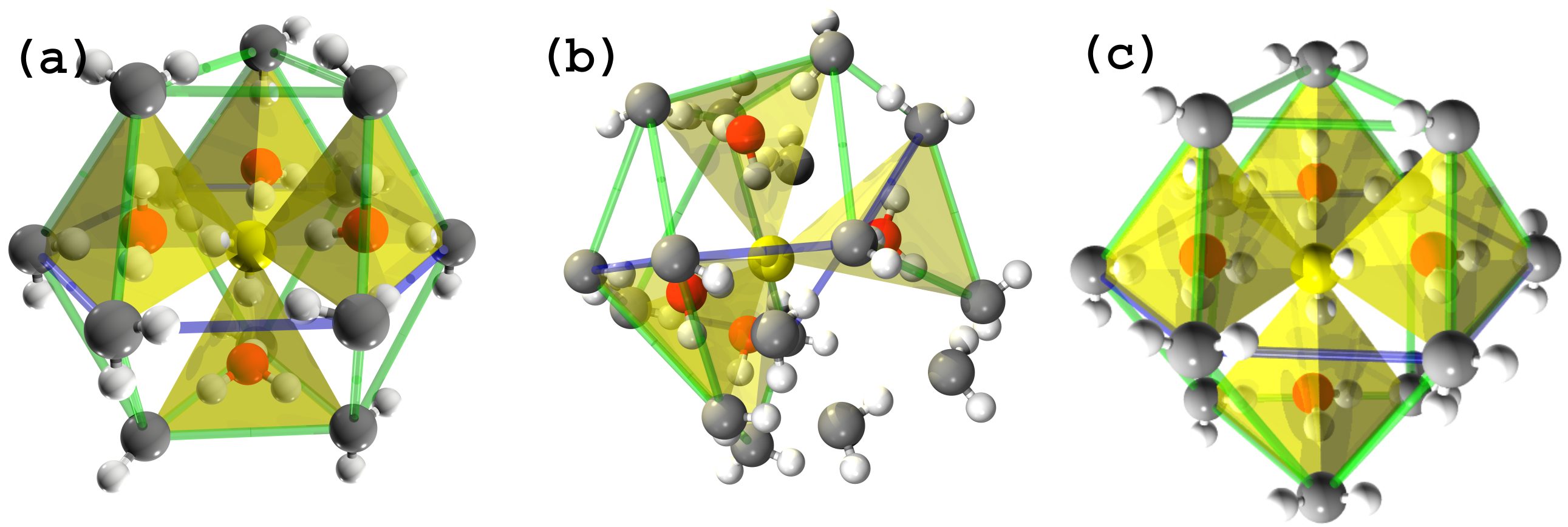}
    \caption{(a) Local structure of I$h$, (b) LDA, and (c) ice I$c$. The central water molecule is depicted in yellow and is located 
             at the shared vertex of the four tetrahedra depicted in yellow. The water molecules in the first shell of the 
             central (yellow) water molecule are shown in red and are located at the center of the four (yellow) tetrahedra. 
             The water molecules on the second shell of the central (yellow) molecule are depicted in gray and are located at the outer 
             vertices of the (yellow) tetrahedra. In the case of I$h$, the gray molecules 
             describe an anticuboctahedron whose structure is emphasized 
             by the green and blue lines. In the case of LDA, one of the four (yellow) tetrahedra of the anticuboctahedral cage 
             is broken. In the case of I$c$, the second shell of neighbors describes the Archimedean solid cuboctahedron.}
 \label{fig:FigIhLDA}
\end{figure}

A comparison of Figs.~\ref{fig:FigSth} and ~\ref{fig:scores} (b) shows that, although LDA and I$h$ have very similar tetrahedrality 
(and density), they differ remarkably at the level of the second shell. 
This is shown in Fig.~\ref{fig:FigIhLDA} (a) and ~\ref{fig:FigIhLDA} (b) that include, respectively, a 
representative arrangement of water molecules in the first and second hydration shell of I$h$ and LDA, 
taken from our MD simulations.
The yellow water molecule is the shared vertex of four tetrahedra (emphasized in yellow), that are the source of the 
high tetrahedrality of both samples. The centers of each of these tetrahedra are occupied by the red water molecules;
these four (red) water molecules constitute the first hydration shell of the central (yellow) molecule.
The outer vertices of the (yellow) tetrehedra are occupied by 12 molecules, shown in grey. These molecules constitute the 
second hydration shell of the central (yellow) water molecule.
The oxygens in the second shell of I$h$ describe an anticuboctahedron, emphasized by the blue snapshot in 
Fig.~\ref{fig:scores} (b). 
A comparison of Figs.~\ref{fig:FigIhLDA} (a) and (b) shows that, in the local structure of LDA, one of the 
four tetrahedra is broken, leading to the opening of the anticuboctahedral cage characteristic of I$h$.
For comparison, included in Fig.~\ref{fig:FigIhLDA} (c) is the local structure of ice I$c$. 
The second shell of neighbors of I$c$ (gray O atoms) describes the Archimedean solid 
cuboctahedron~\cite{archimedean} and differs from the 
anticuboctahedron in the local structure of ice I$h$. Specifically, the anticuboctahedron (I$h$) is the $27^{th}$ 
Johnson solid~\cite{Johnson} and differs from the cuboctahedron (I$c$) by a rotation of $120^{\circ}$ 
in one of the four (yellow) tetrahedra (Fig.~\ref{fig:FigIhLDA} (c)). 
\subsection{Clustering analysis}\label{cluster}
\begin{figure}
 \centering
    \includegraphics[scale=.33]{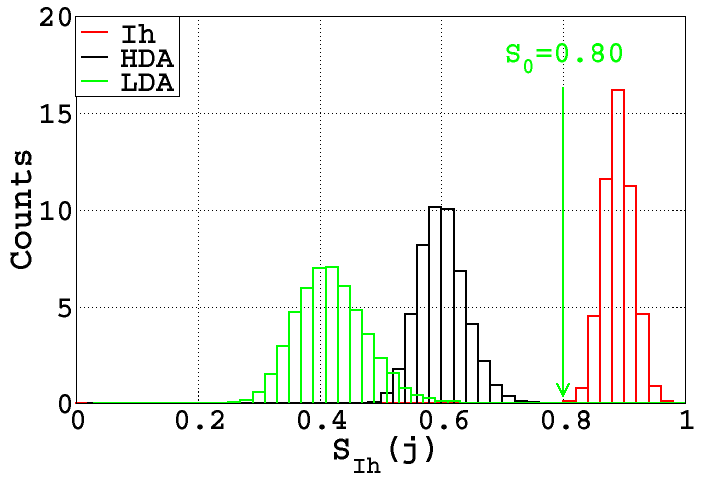}
    \caption{Histogram of the LOM $S_{Ih}(j)$ (eq.~\ref{eq:Eq1}) calculated using, as a reference, the second shell of neighbors of ice I$h$ for
            HDA (black) at p$=3.0$ GPa, I$h$ (red) at p$=0.01$ GPa, and for LDA (green) at p$=0.01$ GPa, respectively.}
 \label{fig:overlap}
\end{figure}

In this section, we describe the structural changes underlying the I$h$-to-HDA and the LDA-to-HDA transformations
by looking at the spatial distribution of ice IV-, ice I$h$-, LDA-, and HDA-like water molecules through the system. 
We classify a water molecule $j$ as I$h$-like molecule if the corresponding LOM $S_{Ih}(j)>S_0$, and
as ice IV-like molecule if its LOM $S_{IV}(j)>S_0$; otherwise, the molecule is considered to belong to an amorphous ice. 
The value $S_0=0.8$ is chosen because 
for ices $\alpha=$ I$h$, IV, $S_{\alpha}(j)>0.8$, while $S_{\alpha}(j)<0.8$ for molecules in LDA and HDA; see 
Figs.~\ref{fig:overlap} and ~\ref{fig:hist} (c).
In order to distinguish between LDA- and HDA-like molecules, we consider the LOM $S_{IV}(j)$. As shown in 
Fig.~\ref{fig:hist} (c), molecules in LDA (HDA) are characterized by approximately $S_{IV}(j)<S_{0'}$ ($S_{IV}(j)>S_{0'}$) 
with $S_{0'}=0.58$. \par
This classification of water molecules allows us to identify clusters of LDA, HDA, ice I$h$, and ice IV. Specifically, two molecules, 
of the same kind (LDA, HDA, I$h$, and IV), are considered to belong to the same cluster if they form a HB. In this work, 
we consider that two water molecules form a HB if a H atom of one of these molecules is within a distance $d_{OH}=2.2$ \AA 
from the 
O atom of the other molecule; this value of $d_{OH}$ corresponds to the location of the first minimum in the OH radial distribution 
function of glassy water at $p=3$ GPa (HDA)~\cite{martelli_prep}.
With respect to other HB definitions,
this definition leads to a fully formed tetrahedral network in LDA and HDA~\cite{martelli_prep} which accounts for the large-scale 
properties of both amorphous ices~\cite{martelli_hyperuniformity}, in agreement with experimental results~\cite{lin_2018}.
\subsubsection{The I$h$-to-HDA transformation}
\begin{figure}
 \centering
    \includegraphics[scale=.40]{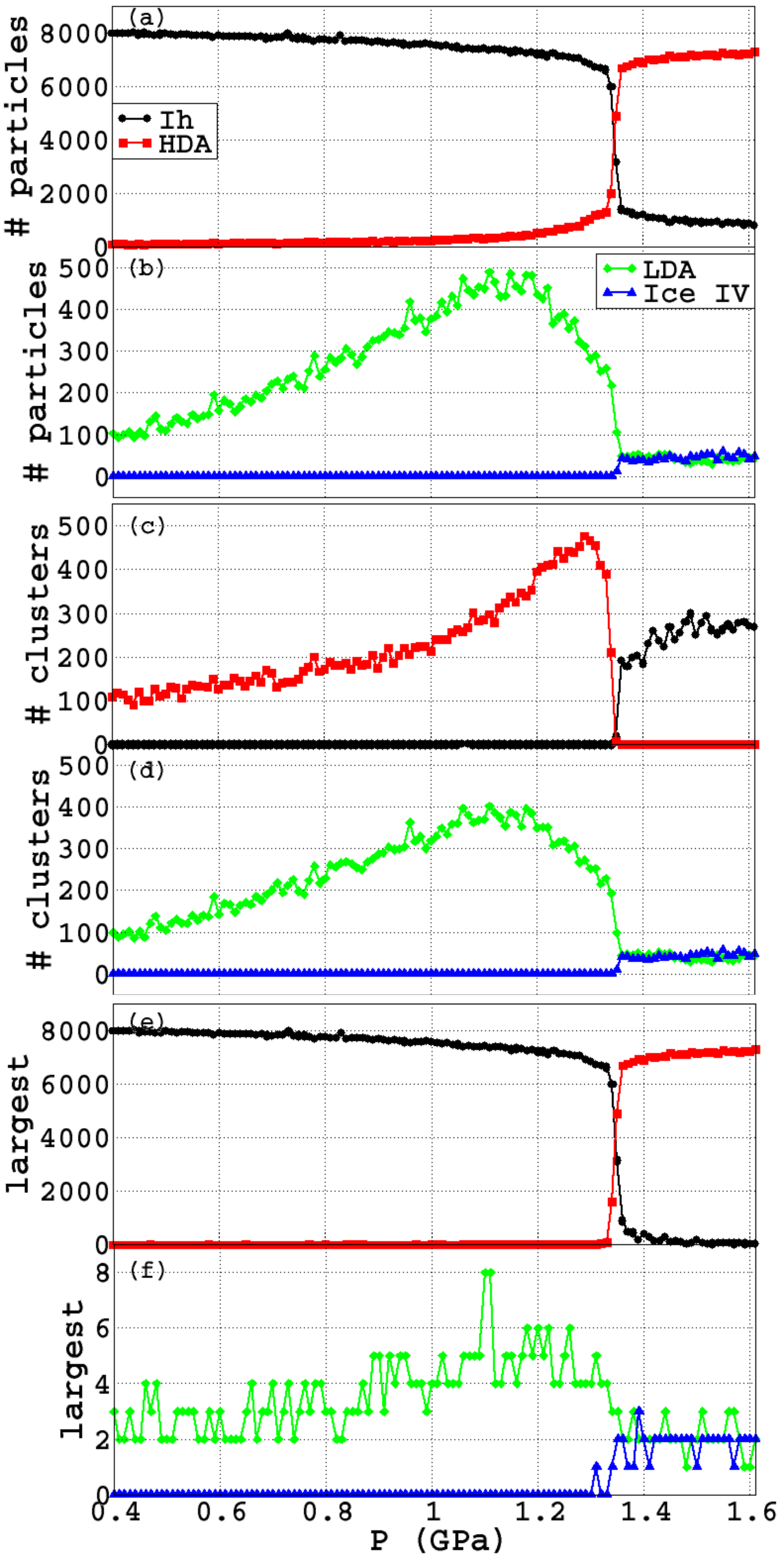}
    \caption{(a) Number of I$h$- (black circles) and HDA-like (red squares) molecules, and (b) LDA- (green diamonds) and ice 
             IV-like (blue triangles) molecules during the I$h$-to-HDA transformation. The corresponding number of clusters are 
             indicated in (c) and (d).  (e) and (f) show the number of molecules composing the ice I$h$-, ice IV-, LDA-, 
             and HDA-clusters.
             In (c), both the number of I$h$ clusters below the transition pressure and the number of 
             HDA clusters above the transition pressure are equal to $1$.}  
 \label{fig:clusters_Ih}
\end{figure}

Fig.~\ref{fig:clusters_Ih} (a) shows the number of I$h$-like (black circles) and HDA-like molecules (red squares) during the 
I$h$-to-HDA transformation for the pressure window $0.4\leq p \leq1.6$ GPa. Similarly, Fig.~\ref{fig:clusters_Ih} (b) shows 
the number of LDA-like molecules (green diamonds) and ice-IV-like molecules (blue triangles) during the same transformation. 

At $p=0.6$ GPa, the number of I$h$-like molecules is large, $\approx 8000 \approx N$, and decreases smoothly upon further compression 
to $p=1.34$ GPa. As shown in Figs.~\ref{fig:clusters_Ih} (a) and ~\ref{fig:clusters_Ih} (b), the decrease in I$h$-like molecules 
is accompanied by an increase in HDA-like 
molecules, from $\approx 0$ at 0.6 GPa to $\approx 1500$ at $p=1.34$ GPa. We note that, at these pressures, the number of 
LDA-like molecules is relatively small ($<500$) and it reaches a maximum at $p=1.1-1.2$ GPa, while there are no ice IV-like 
molecules (at $p<1.34$ GPa). In correspondence with the I$h$-to-HDA transformation at $p=1.35$ GPa, we observe a significant drop 
in the number of I$h$-like molecules, from $\approx 6700$ to $\approx  1500$. which is accompanied by an increase in the number 
of HDA-like molecules, from $\approx 1500$ to $\approx  6700$. Interestingly, at $p>1.35$ GPa, HDA contains a small number of 
residual, LDA and IV molecules; e.g., at $p=1.4$ GPa, there are only $\approx 50$ molecules of LDA and ice IV.
In addition, depending on pressure, specifically HDA$_{Ih}$ may also contain non-negligible amounts of  residual ice I$h$ molecules. 
At $p=1.6$ GPa, there are $\sim1000$ I$h$-like molecules (arranged in $\sim450$ clusters) while ice I$h$ molecules are 
absent at $p=3$ GPa.\par
Figures.~\ref{fig:clusters_Ih} (c) and ~\ref{fig:clusters_Ih} (d) show the number of clusters associated to LDA, HDA, ice I$h$ and IV.  
At pressures below the transformation pressure, the system contains only one I$h$-cluster with several, 
small disconnected HDA-like and LDA-like clusters. Indeed, the largest LDA- and HDA-cluster are composed 
by less than 10 molecules (at $p<1.34$ GPa) (Fig.\ref{fig:clusters_Ih} (f) and ~\ref{fig:clusters_Ih} (e)).
On the other hand, in correspondence with the I$h$-to-HDA transformation, the number of I$h$-like clusters increases, 
from $1$ to $\approx 200-250$, while the number of HDA-clusters reduces to $1$.
As shown in Fig.~\ref{fig:clusters_Ih} (e), the largest HDA-cluster has $\approx 7000$ molecules and hence, it includes most of 
the molecules in the system while, instead, the largest I$h$-cluster is small ($<10$ molecules in size) and therefore, they 
are within the HDA matrix.
It follows that the I$h$-to-HDA transformation is reminiscent of a transformation driven by a
spinodal decomposition, rather than by a nucleation-and-growth process as found during crystallization.
\subsubsection{The LDA-to-HDA transformation}
\begin{figure}
 \centering
    \includegraphics[scale=.33]{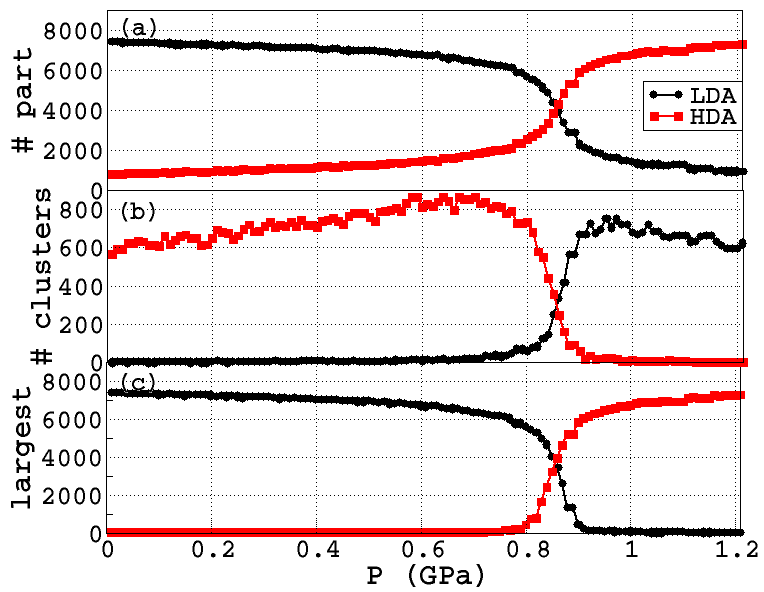}
    \caption{(a): Number of LDA and HDA clusters in black and red, respectively for the LDA-to-HDA transformation. 
     (b): Largest LDA and HDA clusters.}
 \label{fig:clusters_LDA}
\end{figure}

A similar picture holds for the LDA-to-HDA transformation. However, in this case, the transformation is
smoother and there are no I$h$-like molecules at any pressure. Accordingly, we only discuss the roles of LDA and
HDA during the LDA-to-HDA transformation. \\
Fig.~\ref{fig:clusters_LDA} (a) shows the number of LDA-like (black circles) and HDA-like molecules (red squares) during the 
LDA-to-HDA transformation for the pressure window $0.01\leq p \leq1.2$ GPa. At $p=0.01$ GPa, the number of LDA-like molecules is large, $\approx 7500$, and decreases 
smoothly upon further compression 
to $p=0.8$ GPa. As shown in Figs.~\ref{fig:clusters_LDA} (a), the decrease in LDA-like molecules 
is accompanied by an increase in HDA-like 
molecules, from $\approx 600$ at $p\sim0$ GPa to $\approx 2000$ at $p=0.80$ GPa. 
In correspondence with the onset of the LDA-to-HDA transformation at $p=0.80$ GPa, we observe a significant drop 
in the number of LDA-like molecules, from $\approx 6000$ to $\approx  2000$ at $p=0.9$ GPa, which is accompanied by an 
increase in the number of HDA-like molecules, from $\approx 2000$ to $\approx 6000$. \par
Figure~\ref{fig:clusters_LDA} (b) shows the number of clusters associated to LDA and HDA.
At pressures below the transformation pressure, the system contains one LDA-cluster with several, 
small disconnected HDA-like clusters that increase in number upon compression, reaching a maximum of $\sim800$ clusters at 
$p\sim0.7$ GPa. The largest HDA-cluster at $p < 0.8$ GPa is composed by less than $10$ molecules; see Fig.~\ref{fig:clusters_LDA} (c).
On the other hand, in correspondence with the pressure window $0.8\lesssim p\lesssim 1.0$ GPa, the number of LDA-like clusters 
increases, from $\approx 100$ to $\approx 700$ and they decrease at higher pressures, while the number of HDA-clusters reduces to one.
At $p>0.9$ GPa, this HDA-cluster has $N\sim8000$ molecules and hence, it includes most of 
the molecules in the system while, instead, the largest LDA-cluster is small ($<10$ molecules in size) and therefore, they 
are within the HDA matrix.
It follows that, in analogy with the I$h$-to-HDA transformation, the LDA-to-HDA transformation is reminiscent of a 
transformation driven by spinodal decomposition.
\section{Summary and conclusions}\label{Conclusions}
We performed classical MD simulations and explored in detail the structure of water at short and intermediate scales 
during the LDA-to-HDA and I$h$-to-HDA transformations.  Specifically, 
by using the LOM developed in Ref.~\cite{martelli_LOM}, we searched for locally ordered crystalline domains 
composed by the first and/or second shell of neighbors in 
the amorphous ices formed during the I$h$-to-HDA and the LDA-to-HDA pressure-induced transformations. Our results confirm that 
LDA and HDA are indeed amorphous, i.e., they lack of polydispersed ice-like structures. Surprisingly, we find that HDA 
contains a small number of domains that are reminiscent of ice IV. The presence of ice IV-like domains provides some 
support to the hypothesis that HDA could be a 'derailed' state along the I$h$-to-ice IV pathway~\cite{shephard_HDA}.  However, the 
HBN of these domains differ from the HBN of ice IV.  The ice IV basic structure includes the interpenetration of two HBN 
(Fig.~\ref{fig:ice4}). However, interpenetration of HBNs is highly improbable in amorphous structures such as HDA since it 
constrains the local possible orientations of water molecules and hence, it tends to reduce the entropy of the amorphous ice. 
In our view, the transformation of HDA to ice IV is not fully understood and requires further investigation.\par
By characterizing the structure of water at the first and second hydration shells, we also provided some understanding of 
the differences and similarities in the molecular changes underlying the LDA-to-HDA and I$h$-to-HDA transformations. 
Our results indicate that both nonequilibrium transformations are two-steps processes where, first, a small distortion of the HBN 
occurs at low pressures and then, a sudden, extensive re-arrangement of HBs at the corresponding transformation pressure occurs. 
The I$h$-to-HDA and LDA-to-HDA transformations occur when LDA and I$h$ have similar local order, as quantified by various 
score functions (e.g., $S_{IV}$, $S_{Ih}$, and $S_{th}$). Since I$h$ has a perfect tetrahedral HBN, while LDA does not, it 
follows that higher pressures are needed to distort the HBN of Ih relative to LDA. Accordingly, Ih can resist larger 
pressures than LDA before collapsing to HDA, as found in experiments~\cite{johari_2000,johari_2004}.\par
From the microscopic point of view, I$h$ and LDA are structurally similar at small lenght-scales,
both being characterized by a nearly perfect regular tetrahedral HBN. Yet, the I$h$-to-HDA transformation is sharp 
while the LDA-to-HDA transformation is gradual. Our results suggest that the gradualness of the LDA-to-HDA transformation is due to 
different structural configurations occurring at the level of the second shell of neighbors: I$h$ has a well-formed anticuboctahedral 
structure and an ordered HBN (which breaks rapidly upon compression), while LDA has an open cage that causes some degree of 
disorder in the HBN (Fig.~\ref{fig:FigIhLDA}) (and breaks less rapidly upon compression). \par
We also find that both I$h$-to-HDA and LDA-to-HDA transformations are first-order-like phase transitions occurring via spinodal 
decomposition in which HDA clusters percolate within the I$h$/LDA sample. At pressures below the I$h$/LDA-to-HDA transformations, 
the system consists of a single large cluster of I$h$/LDA, composed by most of the molecules in the system, plus  
HDA clusters, composed of few water molecules. At pressures above the transformation pressure, the roles of HDA and I$h$/LDA 
are inverted, i.e., the system is composed of a single HDA-cluster composed by most of the molecules in the system, and 
I$h$/LDA clusters (depending on whether the starting phase is I$h$ or LDA) that are composed by a few 
water molecules.\par
We conclude by noting that HDA${_{Ih}}$ and HDA${_{LDA}}$, at $p=3$ GPa, are practically identical in terms of water structure at 
the first- and second-hydration shells. However, minor differences in structure seem to exist at lower pressures, closer to 
the collapse of I$h$ and LDA to HDA. For example, it follows from Fig.~\ref{fig:clusters_Ih}(a) that HDA${_{Ih}}$ contains a few, 
non-negligible number of I$h$-like molecules at $p\sim1.4-1.6$ GPa while, not surprisingly, HDA${_{LDA}}$ does not. This is 
consistent with the picture proposed in Refs.~\cite{loerting_the_2015,aman_colloquium_2016} where uHDA (at $p<1.5$ GPa) is 
considered to contain I$h$ crystallites while other HDA forms, such as eHDA, do not. In our case, HDA${_{Ih}}$ is prepared 
following the same protocol followed in experiments to prepare uHDA and, accordingly, it contains ice I$h$ at $p\sim1.5$ GPa.
Instead, HDA${_{LDA}}$ does not contain ice I$h$, as is the case of eHDA. In the pressure range
$1.4\lesssim p \lesssim 1.6$ GPa, the I$h$-clusters in HDA${_{Ih}}$ are small, composed by $<5-10$ molecules.\par
Our results shed light on the debated structural properties of amorphous ices and indicate that the kinetics 
of the I$h$-to-HDA and LDA-to-HDA transformations requires an in depth inspection of the underlying HBN. Such investigation is 
currently ongoing and will be the subject of a forthcoming publication~\cite{martelli_prep}.

\begin{acknowledgments}  
This work was supported by the STFC Hartree Centre's Innovation Return on Research programme, funded by the Department for 
Business, Energy \& Industrial Strategy.
This work was partially supported by the National Science
Foundation (CBS-1604504 award to NG) and a grant of computer time from
the City University of New York High Performance Computing Center under NSF Grants CNS-0855217, CNS-0958379 and ACI-1126113.
R. C. acnowledges partial support from the U.S. Department of Energy under award No. DE-SC0008626.
S. T. was supported by the National Science Foundation under Award No. DMR1714722.
\end{acknowledgments} 

\bibliographystyle{unsrt}
\bibliography{References}

\end{document}